# An off-axis, wide-field, diffraction-limited, reflective Schmidt Telescope

Will Saunders*

Anglo-Australian Observatory, PO Box 296 Epping, NSW 1710, Australia

## ABSTRACT

Off-axis telescopes with unobstructed pupils offer great advantages in terms of emissivity, throughput, and diffraction-limited energy concentration. For most telescope designs, implementation of an off-axis configuration imposes enormous penalties in terms of cost, optical difficulty and performance, and for this reason off-axis telescopes are rarely constructed. However, for the reflective Schmidt design, implementation of an off-axis configuration is very straightforward, and involves only a modest optical penalty. Moreover, the reflective Schmidt gets particular benefits, avoiding the obstruction of its large focal plane and support column, and gaining a highly accessible, gravity-invariant prime focus, capable of accommodating very large instrumentation. We present an off-axis f/8 reflective Schmidt design for the proposed 'KDUST' Chinese infrared telescope at Dome A on the Antarctic plateau, which offers simultaneous diffraction-limited NIR imaging over 1°, and close to diffraction-limited imaging out to 2° for fibre-fed NIR spectroscopy.

## 1   INTRODUCTION

Large astronomical telescopes are almost invariably reflective, on-axis, axisymmetric designs, having a substantial structure at the top end, within the telescope beam, containing either a prime focus or a secondary mirror. This arrangement is convenient from an engineering and optical design viewpoint, but suffers from major disadvantages for astronomical use. The most obvious ones are that (a) the primary mirror is vignetted by the top end and its support; (b) these structures increase the NIR emissivity of the telescope, (c) the diffraction pattern arising from the vignetted pupil has reduced energy in the core, and (d) the top end support structure introduces large diffraction spikes into the images – to the extent that they define our mental image of stars as seen by telescopes.

These problems affect radio telescopes even more severely (because the images are always diffraction-limited), but the optics are invariably much easier than in the optical (because the optics are fewer wavelengths across), so going off-axis is much more feasible. For these reason, radio telescopes are now often off-axis (e.g. Owens Valley, Greenbank). Solar telescopes also have major disadvantages from being on-axis, in this case because of heat loads, and huge efforts are now going into off-axis solar telescopes such as ATST[1] and NST[2].

However, the difficulty of designing, producing and mounting off-axis optics for wide-field astronomical telescopes are so formidable, that this has never been done for any major telescope, although various designs have been proposed[3,4]. Compared with an on-axis design of the same collecting area and focal length, the amount of glass to be polished away from a spherical surface, the difficulty of polishing the final figure (given by the difference between radial and sagittal curvatures), and the aberrations, are all increased by an order of magnitude. Essentially, the image-quality, figuring difficulty, and engineering issues involved are all comparable to those of building the fully-filled aperture telescope with the same surface shapes – i.e. a telescope of over twice the aperture and twice the speed.

will@aaoepp.aao.gov.au; http://www.aao.gov.au;
phone +61 2 9372 4800; fax +61 2 9372 4860

Telescopes with only flat or spherical surfaces have an enormous intrinsic advantage in going to an off-axis design, because they avoid the problem of polishing off-axis aspheric surfaces. Maksutov telescopes have this feature, but require large, thick, and (for off-axis designs) wedged corrector lenses, which preclude their use in large telescopes. However, off-axis Maksutovs are used extensively in spectrograph collimators, such as 2dF.

## 2 OFF-AXIS REFLECTIVE SCHMIDT TELESCOPES

For traditional, cadioptric, Schmidt telescopes, the difficulties of polishing an off-axis corrector plate have meant that off-axis designs are not used in professional astronomy. However the Chinese all-reflective Schmidt design for LAMOST[5] has a corrector plate which is polished flat, with the distortions required to produce the Schmidt corrector form produced by actuators. So this design is an obvious candidate for an off-axis configuration[*].

Moreover, the LAMOST design has some disadvantages which are overcome by an off-axis design. The first is that, as a relatively slow telescope (LAMOST is f/5) with a very wide (5°) field, the focal plane is very large, and hence causes significant obstruction losses. The second is that this large focal plane and associated instrumentation must be held up by a pillar, strong enough to take the weight under compression (though in principal it could be suspended from above). This means both additional obscuration, and a uniquely asymmetric diffraction pattern, with much more energy in the single diffraction spike than for traditional telescopes. The LAMOST telescope itself operates so far from the diffraction limit, that this is not a major issue. However, it would be a significant problem for the WHAT?[6], a variant of the LAMOST design proposed for NIR use in Antarctica. Moving the focal plane out of the beam means that the focal plane, which is already gravity-invariant, becomes highly accessible, and capable of accommodating very large instrumentation. By adding a single fold mirror to bring the beam vertical, the instrumentation can be made genuinely gravity-invariant, and of arbitrary size.

However, it also happens that, uniquely for a reflective Schmidt, the penalty in image quality that comes from going off-axis (effectively doubling the speed of the optics) can be partially offset by moving the centre of rotation of the corrector mirror. This is because by far the biggest contribution to the aberrations in a reflective Schmidt comes from the corrector plate not being at the correct distance from the primary, for most rays, whenever the telescope is used at significant off-axis angle. This error increases as the telescope is used at increasing off-axis angles, but also depends on the physical size of the mirror. For an off-axis configuration, there is an additional degree of freedom to choose the centre of rotation of the corrector mirror, and moving the centre of rotation towards the centre of the off-axis segment is clearly beneficial. It turns out empirically that the optimal position for the centre of rotation is about in the centre of the off-axis segment, which is convenient from an engineering point of view. So, compared with an on-axis design of the same focal length, the maximum error in the axial position of the corrector plate is halved by going to an off-axis design.

To quantify this, suppose a reflective Schmidt telescope, with pupil diameter $D$, is used at an off-axis angle of $\eta$ between the centre of the field on the sky and the telescope axis, to observe a field of radius $\theta$ on the sky. Then rays from the edge of the field furthest from the telescope axis, and passing through the edge of the pupil, hit the corrector plate at the wrong radius, as compared with the classic, on-axis, Schmidt value, as shown in Figure 1.

---

[*] However, it introduces terrible problems with terminology, because the reflective Schmidt design is already 'off-axis' in an *angular* sense, with the centre of the field of view often being tens of degrees from the axis of the telescope. In this paper, 'off-axis angle' is used to describe this property, while 'off-axis use' or 'off-axis design' or 'off-axis configuration' means a lateral displacement between the ray from the centre of the field of view through the centre of the pupil, and the axisymmetric axis of the telescope primary and detector.

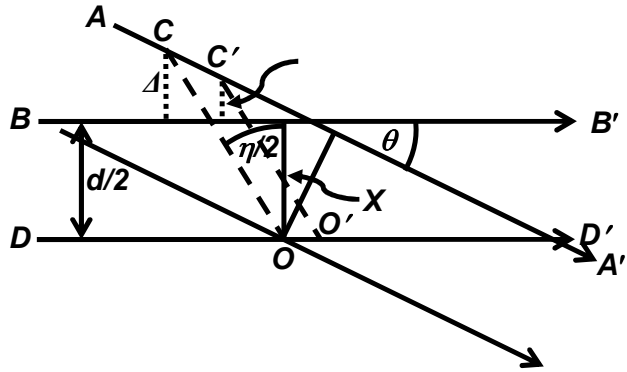

**Figure 1.** Schematic drawing to show the deviation from the ideal location of rays passing through the pupil of a Schmidt with a tilted corrector plate. The picture shows a transmissive rather than reflective tilted corrector plate, for clarity. *O* is the centre of the corrector plate *CO*, which is tilted by an angle $\eta/2$ from its classic orientation (normal to the chief ray *DD′*). *BB′* is a ray from the centre of the field, passing through the corrector plate at a distance from the optical axis of *d*/2. The spherical aberration for this ray can be perfectly corrected by a corrector plate of the correct figure. *AA′* is a ray at the same distance from the centre of the pupil as *BB′* (and so requiring the same spherical aberration correction), but now with a field angle $\theta$. This ray passes through the corrector plate in the wrong place, and the error in the axial position of the ray as it goes through the corrector plate is given by $\Delta$. If the telescope is used in an off-axis configuration, the corrector lens can be moved axially to a new location *C′O′*, reducing the maximum error in the axial position of the ray *AA′* to $\Delta'$.

The size of this deviation is given by

$$\Delta = d/2\,(1 - \cos(\eta/2)/\cos(\eta/2 + \theta)) \approx d/4\,\theta(\eta + \theta)$$

This equation also explains why reflective Schmidts have so much worse image quality than classic cadioptric Schmidt telescopes: while $\theta$ is generally of order a few degrees, $\eta$ (which is zero for cadioptric Schmidts) must be up to 60° or more for a reflective Schmidt, in order to cover the entire sky visible from a temperate site.

For an off-axis configuration, $\Delta$ is greatly reduced, as shown in Figure 1. If the centre of rotation of the corrector plate is placed at the centre of the off-axis corrector plate (marked *X* in Figure 1), then the axial error in the position of *AA′* is halved, in the limit of $\theta \ll \eta$. There is a partially compensating cost, in that rays such as *DD′* now have an error, where they had none before, but as the aberrations increase as the cube of distance off-axis, this error is negligible.

The wavefront error $\Delta\lambda$ deriving from this axial error, for a Schmidt telescope of focal length *f*, and for a ray at radius r from the axis of optical symmetry, varies as

$$\Delta\lambda \propto ((r+\Delta)^4 - r^4)/f^3 \propto \Delta\,r^3/f^3$$

So suppose we compare an off-axis design, with the corrector plate shifted as above, with an on-axis design of the same aperture and focal length. Then *r* is effectively doubled, but $\Delta$ remains the same. There is a further benefit that only a smaller part of the beam has significant axial error, and the final increase in spot size is only a factor ~4. This is the same rate of increase as for e.g., a classic Schmidt or Ritchey-Crétian design, but in this case is not associated with polishing increasingly difficult aspheres.

Of course, the corrector lens does need increased deviation from flat, when moving to the off-axis design. This deviation varies as $d^4/f^3$, so there is a large increase in the displacement, and gradient in the displacement, that must be achieved via the actuators. This will normally require an increase in the number of actuators.

In this paper we present a single aperture, diffraction-limited, $f/8$ design suitable for the proposed Chinese KDUST infrared telescope for Dome A.

## 3   DIFFRACTION-LIMITED, OFF-AXIS F/8 DESIGN

*General features*

This design is proposed for the Chinese 4m KDUST telescope[7], proposed for Dome A on the Antarctic Plateau, at 80°S. The KDUST telescope will be NIR-optimised, to take advantage of the very low sky thermal backgrounds, especially at 2.4μm. At this wavelength, the diffraction-limited image FWHM is 0.14". The image quality is also expected to be superb (once above the ~20m boundary layer); even better than at Dome C, where the median seeing is 0.27-0.36" at 0.5μm[8,9]. Taking into account the competing desires to properly sample the seeing, and to cover as much area as possible for a given detector cost, a suitable plate-scale would be about 0.1"/pixel, or (for the expected 15μm pixels of the HAWAII 4RG detectors), a focal length of 31m. The diameter of the pupil is fairly arbitrarily fixed at 4m, the proposed original diameter for KDUST, giving an f/8 telescope.

The general proposed layout is much as for WHAT?. The main optical axis is horizontal, with spherical primary mirror vertical, and the reflective Schmidt corrector plate operating also as a sidereostat. At Dome A it would be tempting to tilt the entire telescope by 10° to make it equatorial: this means the corrector mirror has to track in azimuth only, and also fixes the pupil for each observation. Image derotation is still required, but the whole telescope, when in use, would then have only two moving mechanisms, plus the actuators for maintaining the corrector plate shape.

To move to an off-axis design just involves tilting the primary mirror, to put the focal plane either below or to one side of the primary beam. Tilting it about a vertical axis gives a somewhat better image quality, so we adopt this as the design. The field of view adopted is 1° for the central, field-flattened portion for imaging. The field-flattened focal plane is 273mm across, enough to ensure that the detector array will be limited by cost rather than by available area. There is also an annular area around it for fibre-fed spectroscopy on a curved focal surface, giving a 2° field overall.

*Optical layout and image quality*

Figure 2 below shows the layout and simultaneous image quality for both flat-fielded 1° and curved 2° fields, when the telescope is used at a typical 30° off-axis angle.

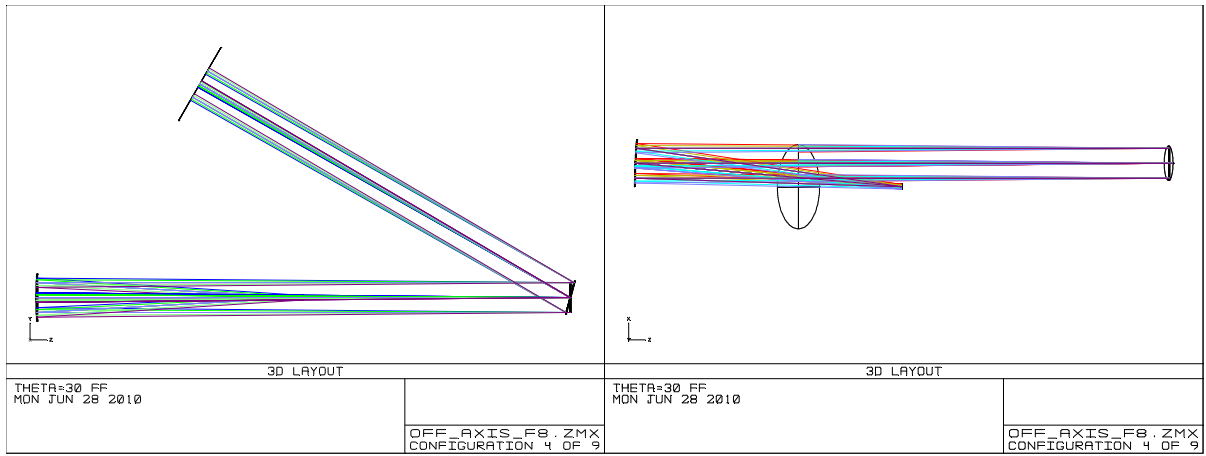

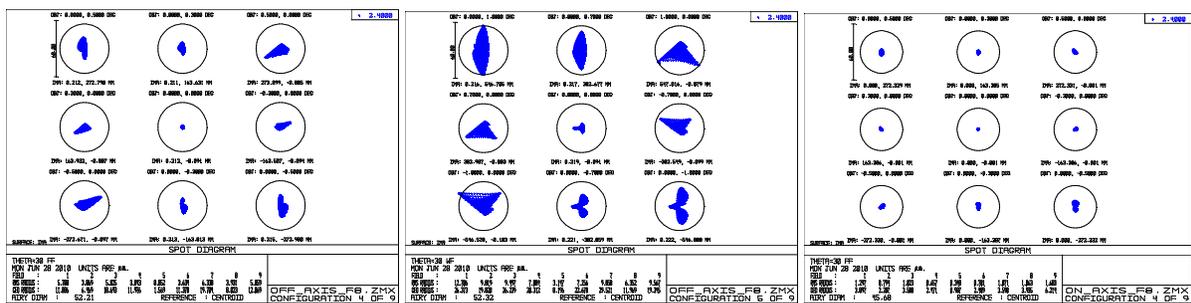

**Figure 2. Telescope layout as seen from (a) aside and (b) above, when used on the meridian at a 30° off-axis angle; (c) image quality for flat-fielded 1° field-of-view showing 60μm scale bar and Airy disc at 2.4μm, (d) image quality for curved 2° field-of-view, (e) for comparison, geometric image quality for a purely on-axis design of the same collecting area and focal length.**

For higher resolution work, sampling the diffraction-limited images over smaller fields, a Gregorian secondary (which could be adaptive) could be used, as proposed for WHAT?.

*Ice-formation*

This paper is primarily about optics rather than the practicalities of building telescopes in Antarctica. However, the problems of ice-formation on optical surfaces in the supersaturated atmosphere of an Antarctic winter are inescapable, and must be addressed. The PILOT design study[11] proposed an air-conditioned dome, not a viable solution for this design. However, it seems feasible that a combination of mirror warming (by a few degrees), and vigorous flushing (at >5 m/s) will prevent frost formation, without introducing unacceptable mirror seeing, especially at infrared wavelengths. This solution also allows large temperature differentials between mirror and air temperature to be tolerated.

*Pupil and cold-stopping*

For the WHAT? telescope, Narcissus mirrors[10] were proposed to control extraneous thermal backgrounds. As part of the PILOT program[11], the thermal backgrounds in Antarctic environments were examined more closely, and it became clear that full cold-stopping is both feasible and desirable for wide-field telescopes. A NIR camera for PILOT was designed, using an Offner-relay cold-stop within the camera. The resulting camera size is large, but for this application it always remains horizontal and in a fixed position, with only axial field rotation correction required. By adding a single fold mirror to bring the beam vertical, the camera can be made genuinely gravity-invariant.

The telescope pupil is easiest defined within the camera, to match a circular beam at the position of the corrector. This means the corrector must be oversized, to fill the pupil when used at large off-axis angles. It is proposed that this oversizing be modest – say 10% to allow sources with off-axis angles up to 45° to be observed – and that a choice of cold stops be offered within the camera to allow sources at larger off-axis angles to be observed with the detectors only seeing cold sky. Alternatively, Narcissus mirrors (reflecting the cold camera) could be used to shield the detectors from seeing unwanted sky or snow when the telescope is used at large off-axis angles. There is a small penalty from the emissivity of the warm mirrors themselves, but a gold-coated mirror at Dome A has an emissivity at least an order of magnitude below the sky brightness, even at 2.4µm.

To avoid vignetting from the primary mirror, over the central 1° field-of-view, requires a primary mirror of diameter 5.1m. For the spectroscopic-only field between radii 0.5° and 1°, requires either greater oversizing of the primary to 5.5m, or Narcissus mirrors (reflecting the cold camera) around the primary.

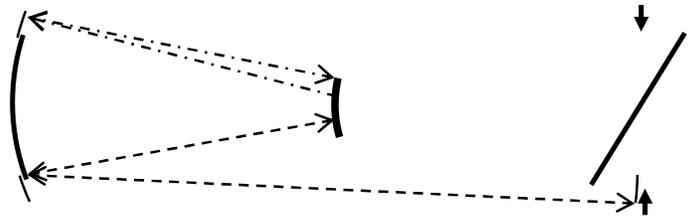

**Figure 3. Schematic indication of how the Narcissus mirrors would be situated. The cold stop within the camera would be conjugate to the nominal telescope pupil shown by the two short arrows. Narcissus mirrors are placed around the corrector mirrors, to shield the focal plane from unwanted radiation when the corrector plate is at large angles (dashed lines). There are also Narcissus mirrors around the primary mirror, to prevent unwanted radiation reaching the outer parts of the focal plane (dot-dashed line).**

*Corrector actuators*

For the corrector plate actuators, the stroke required is a factor 4 larger than for LAMOST, with the longer focal length of the telescope partially offsetting the increased off-axis distance. The displacement gradients are increased by the same factor. This may necessitate more actuators, and thinner segments than for LAMOST.

*Spectroscopy*

Although use at 2.3-2.5µm is the main driver for KDUST, the superb image quality gives a huge benefit for NIR spectroscopy throughout the *JHK* bands, as follows: OH-suppression fibres offer the possibility of dramatic increases in sensitivity for ground-based spectroscopy (and in Antarctica, this could be extended right to the end of the K band at 2.5µm). However, for the foreseeable future, OH-suppression will be an expensive technology, limited by the total number of modes to be suppressed. This number is proportional to the square of both telescope size and seeing. On the other hand, the sensitivity of any telescope observing unresolved sources in the background-limited regime, is proportional to their ratio. So, a telescope only needs to be half as large if placed at a site with seeing twice as sharp, to maintain equal sensitivity. But, the number of modes requiring suppression is reduced by a factor of 16! At f/8, coupling efficiencies into few-mode fibres are excellent[12]. The image quality will be close to diffraction limited much of the time ($D/r_0$ ~1). Starbugs[13] offer a straightforward way of fibre positioning at these temperatures.

## 4   ACKNOWLEDGEMENTS